\documentclass[preprint,authoryear,12pt]{elsarticle}

\usepackage{epsfig}
\usepackage{longtable}

\journal{Radiation Measurements}

\begin{document}
\begin{frontmatter}

\title{CdWO$_4$ crystal scintillators from enriched isotopes for double beta decay experiments}

\author[KINR]{D.V.~Poda\corref{cor1}}
 \ead{poda@kinr.kiev.ua}
 \cortext[cor1]{Corresponding author. Tel.: +38-044-5255283; fax: +38-044-5254463.}
\author[ITEP]{A.S.~Barabash}
\author[INFN-Roma2]{P.~Belli}
\author[INFN-Roma2,Roma2]{R.~Bernabei}
\author[KINR]{R.S.~Boiko}
\author[JINR]{V.B.~Brudanin}
\author[INFN-Roma1,Roma1]{F.~Cappella}
\author[LNGS]{V.~Caracciolo}
\author[LNGS]{S.~Castellano}
\author[LNGS]{R.~Cerulli}
\author[KINR]{D.M.~Chernyak}
\author[KINR]{F.A.~Danevich}
\author[INFN-Roma2,Roma2]{S.~d'Angelo}
\author[KNU]{V.Ya.~Degoda}
\author[LNGS]{M.L.~Di Vacri}
\author[JSC]{A.E.~Dossovitskiy}
\author[NIIC]{E.N.~Galashov}
\author[INFN-Roma1,Roma1]{A.~Incicchitti}
\author[KINR]{V.V.~Kobychev}
\author[ITEP]{S.I.~Konovalov}
\author[KIPT]{G.P.~Kovtun}
\author[LNGS]{M.~Laubenstein}
\author[JSC]{A.L.~Mikhlin}
\author[KINR]{V.M.~Mokina}
\author[KINR]{A.S.~Nikolaiko}
\author[LNGS]{S.~Nisi}
\author[KINR]{R.B.~Podviyanuk}
\author[KINR,INFN-Roma1]{O.G.~Polischuk}
\author[KIPT]{A.P.~Shcherban}
\author[NIIC]{V.N.~Shlegel}
\author[KIPT]{D.A.~Solopikhin}
\author[KINR]{V.I.~Tretyak}
\author[ITEP]{V.I.~Umatov}
\author[NIIC]{Ya.V.~Vasiliev}
\author[KIPT]{V.D.~Virich}

\address[KINR]{Institute for Nuclear Research, MSP 03680 Kyiv, Ukraine}
\address[ITEP]{Institute of Theoretical and Experimental Physics, 117218 Moscow, Russia}
\address[INFN-Roma2]{INFN, sezione Roma ``Tor Vergata'', I-00133 Rome, Italy}
\address[Roma2]{Dipartimento di Fisica, Universit\`a di Roma ``Tor Vergata'', I-00133 Rome, Italy}
\address[JINR]{Joint Institute for Nuclear Research, 141980 Dubna, Russia}
\address[INFN-Roma1]{INFN, sezione Roma ``La Sapienza'', I-00185 Rome, Italy}
\address[Roma1]{Dipartimento di Fisica, Universit\`a di Roma ``La Sapienza'', I-00185 Rome, Italy}
\address[LNGS]{INFN, Laboratori Nazionali del Gran Sasso, I-67100 Assergi (AQ), Italy}
\address[KNU]{Kyiv National Taras Shevchenko University, MSP 03680 Kyiv, Ukraine}
\address[JSC]{Joint Stock Company NeoChem, 117647 Moscow, Russia}
\address[NIIC]{Nikolaev Institute of Inorganic Chemistry, 630090 Novosibirsk, Russia}
\address[KIPT]{National Science Center ``Kharkiv Institute of Physics and Technology'', 61108 Kharkiv, Ukraine}

\begin{abstract}
Cadmium tungstate crystal scintillators enriched in $^{106}$Cd and
$^{116}$Cd were developed. The produced scintillators exhibit good
optical and scintillation properties, and a low level of radioactive
contamination. Experiments to search for double beta decay of
$^{106}$Cd and $^{116}$Cd are in progress at the Gran Sasso
National Laboratories of the INFN (Italy). Prospects to further improve
the radiopurity of the detectors by recrystallization are discussed.
\end{abstract}

\begin{keyword}

CdWO$_4$  crystal scintillator \sep Enriched isotope \sep $^{106}$Cd \sep $^{116}$Cd 
\sep Low counting experiment \sep Radiopurity \sep Double beta decay

\end{keyword}

\end{frontmatter}

\section{Introduction}

Two neutrino double beta ($2\nu2\beta$) decay is the rarest
nuclear transformation ever observed; the half-lives are in the
range of $T_{1/2}\sim10^{18}$--$10^{24}$ yr (see e.g. \citep{Tre95,Tre02,Bar10a}). As regards the neutrinoless mode
($0\nu$) of the decay, a particular analysis of the data collected by the Heidelberg-Moscow 
collaboration, exploiting the $^{76}$Ge isotope, was 
presented in \citep{Kla06}; several experiments are now in progress on the same 
and on other isotopes to further investigate the process.
Published limits on the half-life of this process in various isotopes are at level of 
$\lim T_{1/2}\sim10^{23}$--$10^{25}$ yr (see e.g. \citep{Tre95,Tre02,Bar10b,Gom12}). 
Investigations of the $0\nu2\beta$ decay are related with several
fundamental topics of particle physics: the lepton number
violation, the nature of neutrino (Majorana or Dirac particle), an
absolute scale of neutrino mass and the neutrino mass hierarchy
(see \citep{Rod11,Ell12,Ver12,Gom12} and references therein).

One of the most sensitive 2$\beta$ experiments was realized in the
Solotvina underground laboratory (Ukraine) with the help of
$\approx0.3$ kg radiopure cadmium tungstate scintillators enriched
in $^{116}$Cd ($^{116}$CdWO$_4$) \citep{Dan03}. CdWO$_4$ crystal
scintillators were also successfully used to search for $2\beta$
processes in $^{106}$Cd \citep{Dan96,Dan03} and $^{108,114}$Cd
\citep{Bel08}. The experiments showed that CdWO$_4$ scintillator is
a promising detector thanks to possibility to realize calorimetric
``source = detector'' experiment with a high detection efficiency,
low level of intrinsic radioactivity, good scintillation
properties,  ability of pulse-shape discrimination in order to suppress
background caused by intrinsic radioactive contamination, and long
operation stability. In addition, $^{116}$Cd and $^{106}$Cd
isotopes are favorable candidates for 2$\beta$ experiments thanks
to the high energy of decay ($\approx 2.8$ MeV \citep{Tre95,Tre02}), promising
theoretical estimations (e.g. see in \citep{Tre95,Tre02}), relatively large 
isotopic abundance \citep{Ber11} and
possibility of enrichment by ultra-centrifugation \citep{Art97}.

High quality $^{106}$CdWO$_4$ crystal scintillator enriched in
$^{106}$Cd was developed for the first time \citep{Bel10}. Large
volume $^{116}$CdWO$_4$  crystal was produced to investigate
2$\beta$ decay of $^{116}$Cd \citep{Bar11}. Experiments with these
detectors \citep{Bel12,Bar12} demonstrate their high scintillation
quality and low level of radioactive contamination. Nevertheless,
taking into account that high sensitivity $2\beta$ experiments
require very low, ideally zero, background, an R\&D to further improve
the radiopurity of the scintillators by recrystallization
is in progress. Some preliminary results of this work are
presented here.

\section{Development and characterization of enriched $^{106, 116}$CdWO$_4$ scintillators}

To develop crystal scintillators from enriched material, one needs 
to minimize loss of costly isotopically enriched materials,
maximize yield of crystal scintillators, prevent radioactive
contamination. Taking into account these demands, the production
of the enriched $^{106}$CdWO$_4$ and $^{116}$CdWO$_4$ crystals has
been performed in the following steps: a) purification of the
isotopically enriched cadmium metal samples by means of heating with filtering 
and subsequent distillation through getter filter; 
b) chemical purification and synthesis of cadmium
tungstate compounds; c) crystal growth; d) cut to produce
scintillation element(s). The purity grade of all the materials
was controlled at all the steps by using mass and atomic
absorption spectrometry.

The purification of the enriched samples was performed by distillation 
through graphite filter in vacuum of 10$^{-5}$ Torr \citep{Ber08,Kov11}. 
The procedure allowed to purify the enriched $^{106, 116}$Cd samples to the
level of $\leq$ 1 ppm (Fe, Mg, Mn, Cr, V, Co), $\leq$ 0.2 ppm (Ni,
Cu), and $\leq$ 0.1 ppm (Th, U, Ra, K, Rb, Bi, Pb, Lu, Sm).

After dissolving the metallic cadmium in nitric acid, the purification
of cadmium nitrate was realized by the coprecipitation of the impurities
on a collector. Tungsten was purified by recrystallization of
ammonium para-tungstate. Solutions of cadmium nitrate and ammonium
para-tungstate were mixed and then heated to precipitate cadmium
tungstate:

\begin{center}
 Cd(NO$_3$)$_2$ + (NH$_4$)$_2$WO$_4$ $\to$ CdWO$_4$ + 2NH$_4$NO$_3$.
\end{center}
All the operations were realized by using quartz, Teflon and
polypropylene labware, materials with low level of radioactive
contaminations.

The boules of $^{106}$CdWO$_4$ (with mass of 231 g) \citep{Bel10} and $^{116}$CdWO$_4$ 
(1868 g) \citep{Bar11} crystals (see Fig. 1 (a,b)) were grown in
platinum crucibles by the low-thermal-gradient Czochralski
technique \citep{Pav92}. The total irrecoverable losses of the
enriched cadmium on all the stages did not exceed 3\%.

\nopagebreak
\begin{figure*}[htbp]
\begin{center}
\begin{minipage}[h]{0.19\linewidth}
\center{\includegraphics[width=1\linewidth]{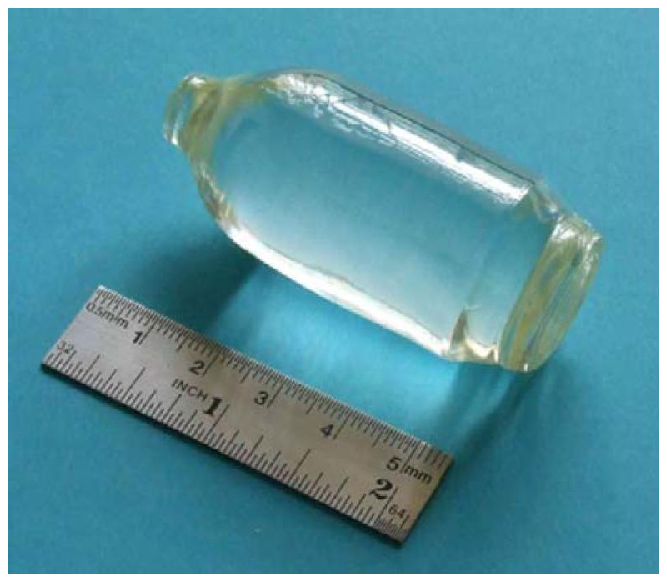}} (a) \\
\end{minipage}
\hspace{5mm}
\begin{minipage}[h]{0.27\linewidth}
\center{\includegraphics[width=1\linewidth]{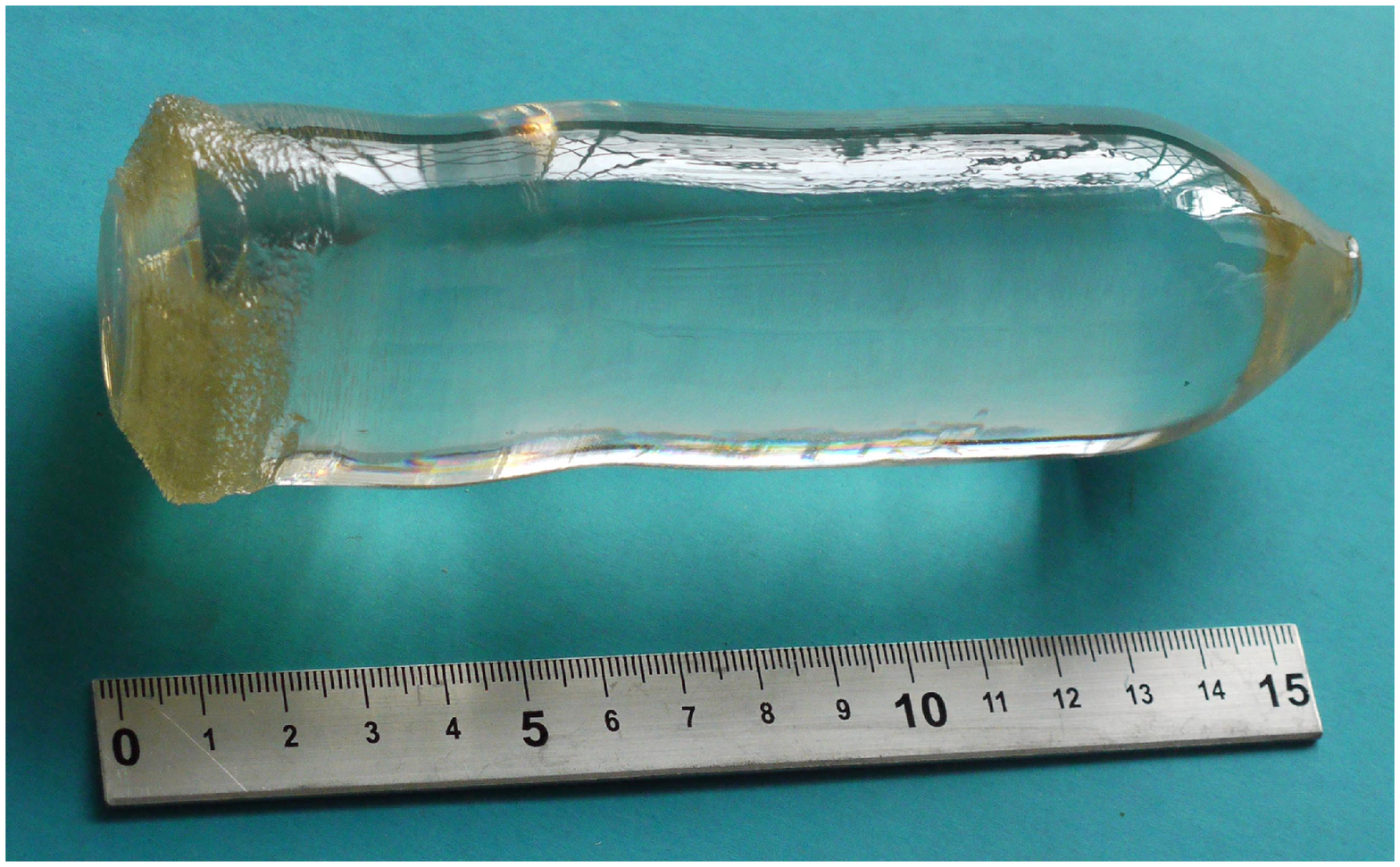}} (b) \\
\end{minipage}
\vfill
\begin{minipage}[h]{0.2\linewidth}
\center{\includegraphics[width=1\linewidth]{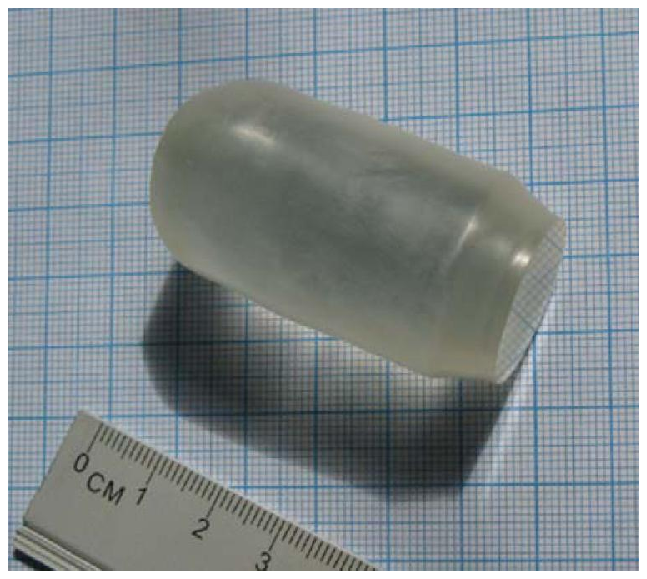}} (c) \\
\end{minipage}
\hspace{5mm}
\begin{minipage}[h]{0.27\linewidth}
\center{\includegraphics[width=1\linewidth]{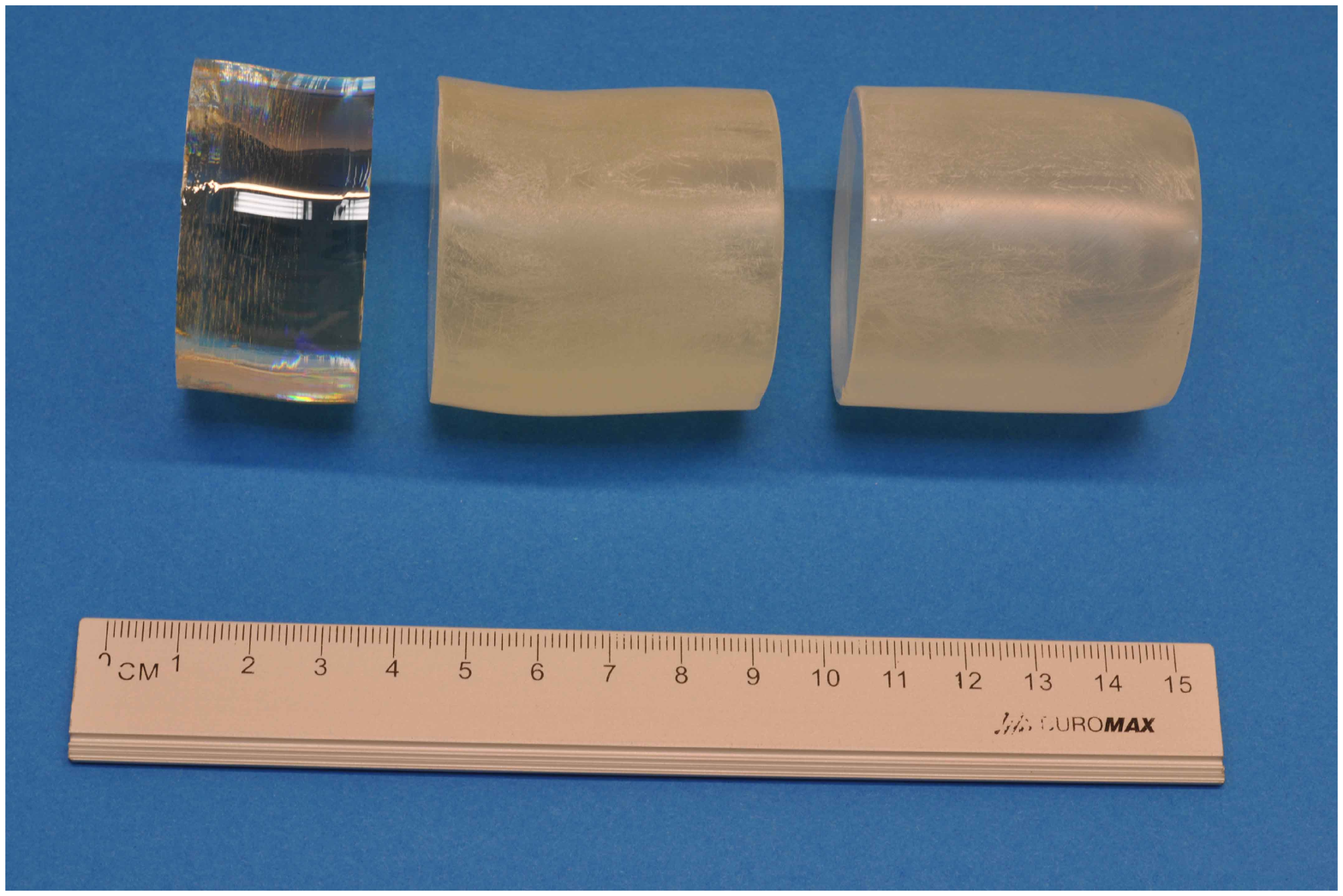}} (d) \\
\end{minipage}
\caption{(Color online) Boules of enriched $^{106}$CdWO$_4$ (a) and 
$^{116}$CdWO$_4$ (b) crystal scintillators. The conic part of the ingots is 
the beginning of the crystals' growth. The scintillation element (c) produced from the 
$^{106}$CdWO$_4$ boule: $\approx\oslash27\times50$~mm, 216 g. 
The $^{116}$CdWO$_4$ crystal samples (d) cut from the boule:
$\approx\oslash46\times25$~mm, 326 g (left); $\approx\oslash45\times46$~mm, 582 g (middle);
$\approx\oslash45\times47$~mm, 580 g (right). The side surface of the $^{106, 116}$CdWO$_4$ 
scintillation elements (except of one) were diffused by using fine-grain polishing paper and 
diamond broach file to improve the uniformity of the scintillation light collection.}
\end{center}
\end{figure*}

One $^{106}$CdWO$_4$ scintillation element (216 g) and three
$^{116}$CdWO$_4$ elements (580 g, 582 g, and 326 g) were produced
from the crystal ingots by cut on the cleavage plane (see 
Fig. 1 (c,d)). Thanks to
the deep purification of starting materials and using of the
low-thermal-gradient Czochralski growing method, the crystals
show excellent optical and scintillation properties. For example,
the attenuation length of the $^{106}$CdWO$_4$ and
$^{116}$CdWO$_4$ scintillation elements were measured as
($60\pm7$) and ($31\pm5$) cm, respectively, at the wavelength of
the emission maximum (490 nm).

The isotopic concentration of the isotopes of interest in the
$^{106}$CdWO$_4$ and $^{116}$CdWO$_4$ crystals was measured 
by mass-spectrometry as 66.41(5)\% and
82.2(1)\%, respectively\footnote{One can compare with isotopic
abundance of $^{106}$Cd and $^{116}$Cd in natural Cd: 1.25(6)\%
and 7.49(18)\%, respectively \citep{Ber11}.}.

The radioactive contamination of the $^{106}$CdWO$_4$ crystal and
of two $^{116}$CdWO$_4$ samples (580 g and 582 g) was measured in
scintillation mode (see Sec. 3). The radioactive contamination of
the 3rd $^{116}$CdWO$_4$ sample (326 g) and the scraps (264 g) after the
crystal growth were checked with the help of ultra-low background
$\gamma$ spectrometry. All the measurements were performed deep
underground at the Gran Sasso National Laboratories of the INFN
(LNGS, Italy). The obtained activities (or limits) are presented
in Table \ref{table:01} where the data on radioactive contamination of enriched
$^{116}$CdWO$_4$ crystals used in the Solotvina experiment and of
CdWO$_4$ scintillators produced from cadmium with the natural
isotopic composition are given for comparison. The developed
$^{106, 116}$CdWO$_4$ scintillators have low level of intrinsic
radioactivity caused by the primordial $^{40}$K and radionuclides
from U/Th chains, anthropogenic $^{90}$Sr and $^{137}$Cs,
cosmogenic $^{110m}$Ag. We explain presence of $\beta$ active
$^{113m}$Cd by strong neutron activation of natural cadmium
samples used in the enrichment procedure. 

The radioactive contamination of the scraps after the $^{116}$CdWO$_4$ 
boule growth is much higher than that of the $^{116}$CdWO$_4$ crystals. 
From the comparison of the data listed in Table 1, one can conclude that 
segregation of K, Th, and especially of Ra, by $^{116}$CdWO$_4$ scintillator 
is very low. It gives a strong signature that
the radioactive contamination of the crystals by $^{40}$K,
$^{228}$Th and $^{226}$Ra can be substantially reduced by
recrystallization.

\begin{table*}[htbp]
\caption{Radioactive contamination of the enriched
$^{106}$CdWO$_4$ \citep{Bel12} and $^{116}$CdWO$_4$ \citep{Bar11}
crystal scintillators, and of the scraps after the
$^{116}$CdWO$_4$ crystals growth. Data of Ref. \citep{Bar11} have
been updated. The upper limits are given at 90\% C.L., 
and the uncertainties of the measured activities at 68\% C.L. 
In the last two columns, data for enriched $^{116}$CdWO$_4$
\citep{Geo96,Dan03} and natural CdWO$_4$
\citep{Geo96,Bur96,Dan96,Bel07} are given for comparison.}
\begin{center}
\resizebox{0.80\textwidth}{!}{
\begin{tabular}{llllllll}

 \hline
 Nuclide    & \multicolumn{7}{c}{Activity (mBq/kg)} \\
\cline{2-8}
(Sub-chain) & $^{106}$CdWO$_4$     & \multicolumn{4}{c}{$^{116}$CdWO$_4$}                 & $^{116}$CdWO$_4$   & CdWO$_4$       \\
\cline{3-6}
 ~           & 216 g               & 580 g       & 582 g       & 326 g       & scraps     & ~ & ~  \\
 \hline
$^{232}$Th  & $\leq0.07$           & $\leq0.08$  & $\leq0.08$  & ~           & ~          & 0.017(5)--0.053(9) & 0.015(8)--$\leq0.1$   \\
$^{228}$Ra  & ~                    & ~           & ~           & $\leq1.3$   & 9(2)       & ~                  & ~                     \\
$^{228}$Th  & 0.042(4)             & 0.041(3)    & 0.067(4)    & $\leq2.0$   & 10(2)        & 0.039(2)           & $\leq0.003$--0.008(4) \\
~           & ~                    & ~           & ~           & ~           & ~          & ~                  & ~                     \\
$^{235}$U   & ~                    & ~           & ~           & $\leq4.0$   & $\leq14$   & $\leq0.007$        & $\leq0.005$           \\
$^{227}$Ac  & ~                    & $\leq0.002$ & $\leq0.002$ & ~           & ~          & 0.0014(9)          & $\leq0.005$--0.014(9) \\
~           & ~                    & ~           & ~           & ~           & ~          & ~                  & ~                     \\
$^{238}$U   & $\leq0.6$            & $\leq0.4$   & $\leq0.6$   & ~           & ~          & $\leq0.6$          & $\leq$(0.045--1.3)     \\
$^{234m}$Pa & ~                    & ~           & ~           & $\leq58$    & $\leq190$  & $\leq0.2$          & ~                     \\
$^{230}$Th  & $\leq0.4$            & $\leq0.06$  & $\leq0.05$  & ~           & ~          & $\leq0.5$          & ~                     \\
$^{226}$Ra  & 0.012(3)             & $\leq0.005$ & $\leq0.005$ & 1.8(8)      & 64(4)      & $\leq$(0.004--0.13)& $\leq$(0.007--0.1)     \\
$^{210}$Po  & $\leq0.2$            & $\leq0.4$   & $\leq0.6$   & ~           & ~          & ~                  & $\leq0.063$           \\
~           & ~                    & ~           & ~           & ~           & ~          & ~                  & ~                     \\
Total $\alpha$ activity  & ~       & ~           & ~           & ~           & ~          & ~                  & ~             	        \\
of Th/U-chains           & 2.1(2)  & 2.10(2)     & 2.93(2)     & ~           & ~          & 1.4(1)--2.8(5)     & 0.26(4)--$\leq0.7$    \\
~           & ~                    & ~           & ~           & ~           & ~          & ~                  & ~                     \\
$^{40}$K    & $\leq1.4$            & $\leq0.9$   & $\leq0.9$   & 18(8)       & 27(11)     & 0.3(1)--$\leq8.8$   & $\leq$(1.7--5)         \\
$^{90}$Sr   &  $\leq0.3$           & ~           & ~           & ~           & ~          & $\leq$(0.2--5.6)   & $\leq$(1--3)           \\
$^{110m}$Ag & $\leq0.06$           & 0.12(4)     & 0.12(4)     & ~           & ~          & ~                  & ~                     \\
$^{113}$Cd  & 182                  & 100(10)     & 100(10)     & ~           & ~          & 91(5)--97          &  558(4)--580(20)      \\
$^{113m}$Cd & $11.6(4)\times 10^3$ & 460(20)     & 460(20)     & ~           & ~          & 1.1(1)--30(10)     & $\leq3.4$--150(10)    \\
$^{137}$Cs  & ~                    & ~           & ~           & 2.1(5)      & $\leq2.1$  & 0.43(6)--1.5(4)    & $\leq$(0.3--0.9)       \\
 \hline

\end{tabular}
 }
 \label{table:01}
 \end{center}
 \end{table*}

\section{Double beta experiments with $^{106}$CdWO$_4$ and $^{116}$CdWO$_4$ scintillation detectors}

Searches for double beta decays of $^{106, 116}$Cd were realized
with the developed $^{106, 116}$CdWO$_4$ scintillators in the low
background DAMA/R\&D set-up installed at the LNGS (Italy)
\citep{Bel12,Bar12}.

Data were accumulated over 6590 h with the $^{106}$CdWO$_4$ and
during 7593 h with the $^{116}$CdWO$_4$ detector. Two neutrino
2$\beta$ decay of $^{116}$Cd was measured with the half-life
$T_{1/2} = (2.5 \pm 0.5) \times 10^{19}$ yr in agreement with the
results of the previous experiments (see \citep{Bar12} and
references therein). Lower half-life limits were set on several
possible $2\beta$ decays in the range of $\lim T_{1/2} \sim
10^{19}$--$10^{21}$ yr for $^{106}$Cd \citep{Bel12} and $\lim
T_{1/2} \sim 10^{21}$--$10^{22}$ yr for $^{116}$Cd \citep{Bar12}, 
all the limits are at 90\% C.L. Most of the limits are stronger than 
those obtained in the previous experiments.

The experiment with the $^{116}$CdWO$_4$ detector is in progress.
A next stage of the experiment with the enriched $^{106}$CdWO$_4$
crystal in coincidence with four HPGe detectors of 225 cm$^3$
volume each (mounted in one cryostat) is in preparation. We
estimate the sensitivity of the experiment, in particular, to
$2\nu\varepsilon\beta^{+}$ decay of $^{106}$Cd, to be at the level
of theoretical predictions, $T_{1/2} \sim 10^{20}$--$10^{22}$ yr.

\section{Conclusions}

Cadmium tungstate crystal scintillators have been developed from
enriched cadmium isotopes: $^{106}$CdWO$_4$ (231 g; 66\% of
$^{106}$Cd) and $^{116}$CdWO$_4$ (1868 g; 82\% of $^{116}$Cd). The
total irrecoverable losses of the enriched cadmium on all the 
stages of scintillators production did not exceed 3\%. The
produced $^{106, 116}$CdWO$_4$ scintillators exhibit excellent
optical and scintillation properties and high level of
radiopurity.

The double beta  experiments using $^{106, 116}$CdWO$_4$ scintillators
are in progress at the Gran Sasso National Laboratories of the
INFN (Italy). The new improved half-life limits on 2$\beta$ decay
of $^{106}$Cd and $^{116}$Cd were set on the level of $\lim
T_{1/2} \sim 10^{19-21}$ yr and $\lim T_{1/2} \sim 10^{21-22}$ yr 
at 90\% C.L., respectively. The 2$\nu$2$\beta$ decay of $^{116}$Cd was observed with
the half-life $T_{1/2}=(2.5 \pm 0.5) \times 10^{19}$ yr, in
agreement with results of previous experiments. The new experiment
to search for 2$\beta$ processes in $^{106}$Cd with the
$^{106}$CdWO$_4$ detector placed in the GeMulti set-up with four
225 cm$^3$ HPGe detectors is in preparation.

We have observed very low segregation of K, Ra, and Th by cadmium
tungstate crystals. It gives a strong signature of possible
improvement of the crystals radiopurity by recrystallization.


\begin{thebibliography}{}

 \bibitem[Artyukhov et. al., 1997]{Art97} Artyukhov, A.A., et. al., 1997. Centrifugal enrichment of cadmium isotopes as the basis for further experiments on physics 
                  of weak interactions, Nucl. Instrum. Methods Phys. Res. A 401, 281--288.
 \bibitem[Barabash, 2010a]{Bar10a} Barabash, A.S., 2010. Precise half-life values for two-neutrino double-$\beta$ decay. Phys. Rev. C 81, 035501, 7 pp.
 \bibitem[Barabash, 2010b]{Bar10b} Barabash, A.S., 2010. Double-beta decay: Present status. Phys. At. Nucl. 73, 162--178.
 \bibitem[Barabash et al., 2011]{Bar11} Barabash, A.S., et al., 2011. Low background detector with enriched $^{116}$CdWO$_4$ crystal scintillators to search
                  for double $\beta$ decay of $^{116}$Cd. J. Instrum. 6, P08011, 24 pp.
 \bibitem[Barabash et al., 2012]{Bar12} Barabash, A.S., et al., 2012. First results of the experiment to search for double beta decay of $^{116}$Cd with the help
                  of enriched $^{116}$CdWO$_4$ crystal. To be published in: Proceedings of the 4th International Conference ``Current Problems
                  in Nuclear Physics and Atomic Energy'' (NPAE-2012),  03--07 September 2012, Kyiv, Ukraine.
 \bibitem[Belli et al., 2007]{Bel07} Belli, P., et al., 2007. Investigation of beta decay of $^{113}$Cd. Phys. Rev. C 76, 064603, 10 pp.
 \bibitem[Belli et al., 2008]{Bel08} Belli, P., et al., 2008. Search for double-$\beta$ decay processes in $^{108}$Cd and $^{114}$Cd with the help of the low 
                  background CdWO$_4$ crystal scintillator. Eur. Phys. J. A 36, 167--170.
 \bibitem[Belli et al., 2010]{Bel10} Belli, P., et al., 2010. Development of enriched $^{106}$CdWO$_4$ crystal scintillators to search for double $\beta$
                  decay processes in $^{106}$Cd, Nucl. Instrum. Methods Phys. Res. A 615, 301--306.
 \bibitem[Belli et al., 2012]{Bel12} Belli, P., et al., 2012. Search for double-$\beta$ decay processes in $^{106}$Cd with the help of a $^{106}$CdWO$_4$
                  crystal scintillator. Phys. Rev. C 85, 044610, 10 pp.
 \bibitem[Berglund and Wieser, 2011]{Ber11} Berglund, M., and Wieser, M.E., 2011. Isotopic compositions of the elements 2009 (IUPAC technical report).
                  Pure Appl. Chem. 83, 397--410.
 \bibitem[Bernabey et al., 2008]{Ber08} Bernabey, R., et al., 2008. Production of high purity Cd and $^{106}$Cd for CdWO$_4$ and $^{106}$CdWO$_4$
                  scintillators [in Russian]. Metallofiz. Nov. Tekhnol. 30, 477--486.
 \bibitem[Burachas et al., 1996]{Bur96} Burachas, S.Ph., et al., 1996. Large volume CdWO$_4$ crystal scintillators. Nucl. Instrum. Methods Phys. Res. A 369, 164--168.
 \bibitem[Danevich et al., 1996]{Dan96} Danevich, F.A., et al., 1996. Investigation of $\beta^{+}\beta^{+}$ and $\beta^{+}$/EC decay of $^{106}$Cd. Z. Phys. A 355, 433--437.
 \bibitem[Danevich et al., 2003]{Dan03} Danevich, F.A., et al., 2003. Search for 2$\beta$ decay of cadmium and tungsten isotopes: final results of the
                  Solotvina experiment. Phys. Rev. C 68, 035501, 12 pp.
 \bibitem[Elliott, 2012]{Ell12} Elliott, S.R., 2012. Recent progress in double beta decay. Mod. Phys. Lett. A 27, 1230009, 16 pp.
 \bibitem[Georgadze et al., 1996]{Geo96} Georgadze, A.Sh., et al., 1996. Evaluation of activities of impurity radionuclides in cadmium tungstate crystals.
                  Instrum. Exp. Tech. 39, 191--198 [Translated from Prib. Tekh. Exp. 2 (1996) 45--51].
 \bibitem[Gomez-Cadenas et al., 2012]{Gom12} Gomez-Cadenas, J.J., Martin-Albo, J., Mezzetto, M., Monrabal, F., Sorel, M., 2012. The search for
  neutrinoless double beta decay. Riv. Nuovo Cimento 35, 29--98.
 \bibitem[Klapdor-Kleingrothaus and Krivosheina, 2006]{Kla06} Klapdor-Kleingrothaus, H.V., Krivosheina, I.V, 2006. The evidence for the observation of  $0\nu\beta\beta$ 
                 decay: The identification of $0\nu\beta\beta$ events from the full spectra. Mod. Phys. Lett. A 21, 1547--1566.
 \bibitem[Kovtun et al., 2011]{Kov11} Kovtun, G.P., et al., 2011. Production of radiopure natural and isotopically enriched cadmium and zinc for low background
                  scintillators. Functional Materials 18, 121--127.
 \bibitem[Pavlyuk et al., 1993]{Pav92} Pavlyuk, A.A., et al., 1993. Low Thermal Gradient technique and method for large oxide crystals growth from melt and flux.
                  In: Proceedings of the APSAM-92 meeting (Asia Pacific Society for Advanced Materials), 26--29 April 1992, Shanghai, China;
                  Institute of Materials Research, Tohoku University, Sendai, Japan, p. 164.
 \bibitem[Rodejohann, 2011]{Rod11} Rodejohann, W., 2011. Neutrino-less double beta decay and particle physics. Int. J. Mod. Phys. E 20, 1833--1930. 
 \bibitem[Tretyak and Zdesenko, 1995]{Tre95} Tretyak, V.I., and Zdesenko, Yu.G., 1995. Tables of double beta 
             decay data. At. Data Nucl. Data Tables 61, 43--90.
 \bibitem[Tretyak and Zdesenko, 2002]{Tre02} Tretyak, V.I., and Zdesenko, Yu.G., 2002. Tables of double beta decay data --- an update. At. Data Nucl. Data Tables 80, 83--116.
 \bibitem[Vergados et al., 2012]{Ver12} Vergados, J.D., Ejiri, H., and \^Simkovic, F., 2012. Theory of neutrinoless double-beta decay. Rep. Prog. Phys. 75, 106301, 52 pp.

\end{thebibliography}
\end{document}